\documentclass{ws-ijmpc}
\usepackage{braket}
\usepackage{bbding}
\usepackage{color}
\usepackage{enumitem}
\usepackage{url}
\usepackage{hyperref}
\usepackage{booktabs}
\usepackage[super]{cite} 
\begin{document}

\markboth{J. Spalding \emph{et al.}}
{Survey-Based Calibration of Social Network Model}

\catchline{}{}{}{}{}

\title{Survey-Based Calibration of the One-Community and Two-Community Social Network Models Used for Testing Singapore's Resilience to Pandemic Lockdown}

\author{Jon Spalding}

\address{Institute of Catastrophe Risk Management, Nanyang Technological University \\
50 Nanyang Avenue, Singapore 639798, Republic of Singapore\\
jonthephysicist@gmail.com}

\author{Bertrand Jayles}

\address{Institute of Catastrophe Risk Management, Nanyang Technological University \\
50 Nanyang Avenue, Singapore 639798\\
bertrand.jayles@gmail.com}

\author{Renate Schubert}

\address{Department of Humanities, Social and Political Sciences, ETH Zurich, Haldeneggsteig 4, 8092 Zurich, Switzerland\\ and \\
Future Resilient Systems, Singapore-ETH Centre, Singapore 138602,
Singapore\\
schubert@econ.gess.ethz.ch}

\author{Siew Ann Cheong\footnote{corresponding author}}

\address{Division of Physics and Applied Physics, School of Physical and Mathematical Sciences, Nanyang Technological University, 21 Nanyang Link,\\
Singapore 637371, Republic of Singapore\\
and\\
Future Resilient Systems, Singapore-ETH Centre, Singapore 138602,
Singapore\\
cheongsa@ntu.edu.sg}

\author{Hans Herrmann}

\address{Departamento de Física, Universidade Federal do Ceará, Caixa Postal 6030, Campus do Pici\\
60455-760 Fortaleza, Ceará, Brazil\\
and\\
PMMH, ESPCI, 7 quai St. Bernard, 75005, Paris, France\\
and\\
Future Resilient Systems, Singapore-ETH Centre, Singapore 138602,
Singapore\\
hans@fisica.ufc.br}

\maketitle

\begin{history}
\received{16 November 2024}
\revised{Day Month Year}
\end{history}

\begin{abstract}
A resilient society is one capable of withstanding and thereafter recovering quickly from large shocks. Brought to the fore by the COVID-19 pandemic of 2020--2022, this social resilience is nevertheless difficult to quantify. In this paper, we measured how quickly the Singapore society recovered from the pandemic, by first modeling it as a dynamic social network governed by three processes: (1) random link addition between strangers; (2) social link addition between individuals with a friend in common; and (3) random link deletion . To calibrate this model, we carried out a survey of a representative sample of $N = 2,057$ residents and non-residents in Singapore between Jul and Sep 2022 to measure the numbers of random and social contacts gained over a fixed duration, as well as the number of contacts lost over the same duration, using phone contacts as proxy for social contacts. Lockdown simulations using the model that fits the survey results best suggest that Singapore would recover from such a disruption after 1--2 months.

\end{abstract}

\ccode{PACS Nos.:}

\section{Introduction}

When an engineered system goes down because of a shock (natural or man-made), teams of engineers and technicians would be deployed to make it operational again. The degradation in performance of the engineered system depends on the severity of the shock, but also on the ability of the system to resist the shock and stay up running. After the shock, the time it takes for the performance to recover to pre-shock levels would depend on the number of teams working on the recovery, how competent these teams are, how well they and the management understand the engineered system, and whether there are bottlenecks in the acquisition of materials and resources needed to bring about the recovery. Together these determine how resilient the engineered system is to various shocks.\cite{Jackson2013,Small2018,Cottam2019} Learning from past shocks, the engineered system can even be made more resilient, although trade-offs are frequently necessary.\cite{Youn2011,Woods2015,Ren2017}

Compared to engineered systems, social systems or communities are less understood. We do not have a clear idea how to measure the performance of a social network, nor do we understand how it resists different types of shocks.\cite{Maguire2007,Keck2013} While we do have emergency response teams from the military, the police, fire fighters, and other emergency service personnel deployed for damage control after disasters strike, they are more like engineering teams: they repair damaged infrastructure, rather than `damaged' social structures. In fact, if the shock does not lead to an irreversible collapse, and we give it enough time, a social system recovers on its own, through mechanisms that are not completely understood. Social resilience is important --- and one might argue more important than infrastructural resilience, and therefore we have started work to understand the former better.\cite{https://doi.org/10.1111/disa.12610,doi:10.1061/NHREFO.NHENG-2034} This was also the motivation behind our recent studies simulating how two social network models recover after pandemic lockdowns. In the first of these studies,\cite{Jayles2022a} we simulated pandemic lockdowns that placed restrictions on social interactions in a homogeneous population of individuals. These social interactions determine the equilibrium state of the social network, and also how quickly the single community recovers. In the second of these studies,\cite{Jayles2022b} we simulated the recovery of the social network under similar pandemic lockdowns, but assumed that the individuals are organized into two communities. The results of the two studies are revealing, and we like to believe that they can inform public policy on pandemics (with appropriate modifications to the models, also for other types of shocks). However, to apply the insights to say the Singapore society, we must at the minimum determine values for the model parameters that best describe Singapore. This is the \emph{calibration problem} that we are concerned with in this paper.

Two main types of models are used by computational social scientists for social simulations: (1) system dynamics (SD) models; and (2) agent-based models (ABM). In a social SD model (e.g., the ``Industrial Dynamics" model introduced by Jay Forrester in the 1960s \cite{forrester1959,forrester1971}), aggregate variables are organized into causal loops, which are then coupled into an overall model of stocks (variables that can accumulate over time) and flows (going from one stock variable to another stock variable, as a representation of the interactions between them) \cite{Richardson1981,Roberts1983,Coyle1996,sterman2000}. Parameters of such models are typically visualized as multiplicative taps that control the flows. To estimate these parameters, one approach would be to set up social experiments involving human subjects, one for each flow parameter, and measure them `directly'.\cite{Biegler1986,Dattner2015,Bach2023} Alternatively, when individuals cannot be experimented upon, but their behaviors can be observed, a second approach would be to identify a few macroscopic variables and measure them as functions of time. Following these measurements, one would then simulate the social SD model with different parameter combinations, and compare the simulated macroscopic variables against the empirical macroscopic variables for each of the parameter combination. The parameter combination whose simulations produce macroscopic variables closest to what were measured in the real world is then accepted as the parameters of the calibrated model.\cite{Coyle1985,Keloharju1989,Oliva2003,Elizondo2019} If the simulations over many parameter combinations are too expensive computationally, it is also possible to start by simulating one or more random set of parameters, compare the simulated macroscopic variables against those from the real world, and then based on these differences update the parameters. Many update strategies using different \emph{optimisation techniques} can be used.\cite{Pedregal2004,Chong2013} 

In a social ABM, heterogeneous individuals are explicitly modeled and simulated. These \emph{agents} must be able to (a) set goals (\emph{proactivity}), (b) respond to changes in the environment (\emph{reactivity}), and (c) interact with other agents (\emph{interactivity}) before the model can be called an ABM \cite{Epstein1996,Gilbert2005}. These features are typically implemented as algorithms, with behavioral parameters deeply embedded within them. Unlike parameters associated with observable flows in SD models, behavioral parameters in ABM are not observable. We can still design careful social experiments to probe these parameters, by taking note of the distribution of choices made by human subjects.\cite{Colasante2017} In general, to calibrate ABMs, we must first identify empirical macroscopic observables that can be easily measured, and compare these against outcomes in the simulations. The parameters are adjusted using various optimization approaches,\cite{Lamperti2018,Platt2020,Zhang2020} until simulations reproduce these macroscopic variables in the least squares sense.

In the first of our recent studies,\cite{Jayles2022a} we simulated a dynamic social network model proposed by Jin \emph{et al.}.\cite{JGN2001} We chose to work with this Jin-Girvan-Newman (JGN) model, which is neither an SD model (individuals are simulated) nor an ABM (because of the lack of proactivity and reactivity), because of its simplicity. In this model, individuals represented as nodes on the network form a homogeneous population. At any given time step, these nodes can have degrees $0 \leq z \leq z^*$. Later, responding to a comment that a real-world society like that in Singapore would consist of multiple interacting communities, we simulated the simplest extension of the JGN model to two communities.\cite{Jayles2022b} The parameters in these models are also behavioral, and cannot be directly observed. However, these can be most directly estimated by asking individuals to self report what they do. This led us to design a survey, and we engaged a survey company to obtain more than two thousand responses from a representative sample of the roughly 5.5 million Singapore residents in the third quarter of 2022. The main purpose of this paper is to describe how we designed the survey, and how we used the survey results to calibrate our JGN models of one and two communities. As far as we can tell, this is the first time a simulation model is calibrated using survey data.

We also expect some of our readers to attempt survey-based calibration of their own social models. Hence, the secondary purpose of this paper is to help readers evaluate whether they should use surveys to calibrate social simulation models in their future studies. Although we mentioned that the self-reported behavioral data is the most direct route to estimating the JGN parameters, this is not easy and certainly not without challenges. For the survey-based calibration of other social models, similar challenges may arise. Therefore, we start in Section \ref{sect:models} by describing the original JGN model, our modifications to make simulating the one-community model easier, and additional parameters we introduced for the two-community model. Then in Section \ref{sect:survey}, we describe the survey that we conducted, the key questions we asked for calibration purposes, and how to use their responses to estimate the model parameters. We also describe other questions that go beyond our models for deeper insights into the resilience of the Singapore communities, and more importantly, the historical context surrounding the survey period. In Section \ref{section:methods}, we summarize the survey results, before describing the simulation mapping approach we used to model parameters that fit the survey results best, and thereafter estimate the additional parameters for the two-community model. We then discuss in Section \ref{sect:discussion} how resilient the Singapore society is as implied by our best-fit model, and the three main challenges we faced using survey data to calibrate the social network models, before concluding in Section \ref{sect:conclusions}.

\section{Models}
\label{sect:models}

In Section \ref{sect:JGN} we review the Jin-Girvan-Newman (JGN) model before presenting our modified version in Section \ref{section:mJGN}. Both are effectively models of a homogeneous community. In Section \ref{sect:twoJGN} we describe how the modified JGN can be further adapted into a two-community model.

\subsection{Jin-Girvan-Newman Model of Dynamic Social Networks}
\label{sect:JGN}

The JGN model proposed by Jin, Girvan, and Newman in 2001 is a minimal model that captures the essence of social network evolution for a fixed population of $N$ nodes.\cite{JGN2001} It has four input parameters: (1) $z^*$, the maximum degree for every node; (2) $r_0$, the rate for creating a random link; (3) $r_1$, the rate for creating a social link; and lastly (4) $\gamma$, the rate for randomly deleting a link. We can simulate this dynamical model starting from any initial state, including a completely disconnected network of $N$ single nodes, or a fully-connected network where each node is connected to every other node. However, it is customary practice to simulate the model starting from a random initial network where the density of links can be fine tuned, so that the network will reach equilibrium in a shorter time.

To simulate a society where individuals meet and form new ties, either with total strangers, or with friends of friends, or lose contact with acquaintances they have not met in a long time, the network is then updated every time step following three rules. First, $r_0\frac{N(N-1)}{2}$ pairs of nodes are randomly selected. For each selected pair $(i,j)$, a new link is added between nodes $i$ and $j$, if they are not already connected, and neither node has reached the maximum degree $z^*$. This ensures that all pairs of nodes are selected with equal probability $r_0$. We call this rule \emph{random link addition}, which is responsible for growing connections between disconnected clusters of nodes. Second, $r_1\frac{\sum_i z_i(z_i-1)}{2}$ nodes are chosen at random, such that the probability of choosing node $i$ (whose degree is $z_i$) is set to be proportional to $z_i(z_i-1)$. For each chosen node, a pair of its neighbors $(j_1, j_2)$ is selected at random and then a new link is created between them if it is possible, i.e., there is no existing link between $j_1$ and $j_2$, and $z_{j_1}, z_{j_2} < z^*$. This ensures that all pairs of links with one common node are selected with equal probability $r_1$. This \emph{social link addition} rule is responsible for creating transitive clusters, and replicates the human tendency to preferentially connect to friends of friends. Finally, we select $\gamma \frac{\sum_i z_i}{2}$ nodes so that node $i$ is picked with probability set to be proportional to $z_i$. We then randomly select one of node $i$'s neighbors and delete the link between them. This \emph{random link deletion} rule replicates the human tendency to forget friends over time, and ensures that all links are picked with equal probability $\gamma$ for deletion.

This model is simple to simulate on a computer due to the small number of parameters and its simple algorithm. More importantly, as shown in the original paper \cite{JGN2001}, this algorithm produces high clustering, a well-known feature of real social networks\cite{Newman2003,oliveira2012overview,opsahl2013triadic,tabassum2018social} that other network models failed to reproduce.

\subsection{Modified Jin-Girvan-Newman Model}
\label{section:mJGN}

To be able to compare the results for networks of different sizes, in Ref.~\citeonline{Jayles2022a} we modified the JGN algorithm described above in three important ways. First, in each of the three steps (random link addition, social link addition, and random link deletion) we do not sample fixed numbers of pairs or triplets of nodes to act on. Instead, we let the continuous parameters $r_0 N$, $r_1 N_m$, and $\gamma N_e$ define the means of three Poisson distributions, and at each time step, sample an integer number $\tilde{\mathcal{R}}_0$ of pairs of nodes to add random links, an integer number $\tilde{\mathcal{R}}_1$ of triplets of nodes to add social links, and an integer number $\tilde{\mathcal{G}}$ of random links to delete. This modification of the JGN model produces additional statistical fluctuations, but it also allows us to tune the parameters $r_0$, $r_1$, and $\gamma$ continuously. More importantly, in the original JGN algorithm if one or more of the parameters $r_0 N$, $r_1 N_m$, $\gamma N_e$ are less than $1$, then the rules they represent will no longer be active, since we can only sample an integer number of nodes, triangles, or links. Therefore, we either have one attempt every time step (when the parameter is greater than 1), or zero attempts every time step (when the parameter is less than 1). Using the original JGN algorithm it is not possible to have one attempt every $10$ or every $100$ time step. With our modified algorithm, we treat $r_0 N$, $r_1 N_m$, $\gamma N_e$ as the means $\lambda$ of Poisson distributions and sample attempts from them. Then even for  $\lambda \ll 1$, we can still get samples of $0, 1, 2, \dots$, with rapidly decreasing probabilities. We believe this is a more reasonable way to simulate the three rules.

Second, we let $r_0 N$ instead of $\frac{r_0}{2}N(N-1)$ be the average number of random pairs of nodes selected for random link addition. This ensures the important properties of the model scale linearly in $N$, i.e., are approximately intensive. For social link addition and random link deletion, we chose $r_1 N_m$ and $\gamma N_e$ to be the average numbers of triplets and pairs sampled from the respective Poisson distributions. Here, $N_m = \frac{1}{2}\sum_i z_i (z_i - 1)$ is the total number of triplets and $N_e = \frac{1}{2}\sum_i z_i$ is the number of links. Therefore, $r_1 N_m$ and $\gamma N_e$ have the same dependence on network size as in the JGN model.

We call this the modified JGN (mJGN) model, or for better contrast against the next model, the \emph{one-community model}. As mentioned, statistical fluctuations over individual simulations are stronger in the one-community model than for the JGN model. This problem is especially serious for time-dependent measurements, such as how the average degree recovers after the end of a lockdown. Therefore, we need many samples for each set of parameters, so that we can obtain averages that are smooth functions of time, along with reasonably small error bars that are the uncertainties in these averages.

\subsection{Two-community Model}
\label{sect:twoJGN}

We also modified the one-community model to accommodate two communities.\cite{Jayles2022b} In this \emph{two-community model}, the three steps of the mJGN algorithm are identical to those above except that the $N$ nodes are now divided into two separate communities. Each community has the same rate coefficients as in the mJGN model, except that the random link addition rate $r_0$ has an intra-community value and a inter-community value. For the $\tilde{\mathcal{R}}_0$ random links that we attempt to add at each time step, a proportion $\alpha$ of them will occur between communities, while a proportion $(1-\alpha)$ will occur within communities. To distinguish the two communities, we will attempt to add $\beta(1 - \alpha)\tilde{\mathcal{R}}_0$  random links within community 1, and $(1 - \beta)(1 - \alpha)\tilde{\mathcal{R}}_0$ will be within community 2. The coefficient $\alpha$ tells us how willing members of different communities are to interact and can be labeled the \emph{inter-community parameter}, while $\beta$ tells us how interactive members are within each community and can be labeled the \emph{intra-community parameter}. When $\beta = 0.5$, the two communities are equivalent, and when $\alpha = 0.5$, the two communities are no longer separated by diminished inter-community interactions. Note that the total link addition attempts at each step, $\tilde{\mathcal{R}}_0$, is the same for the one-community and two-community models because the sum of the coefficients listed above, $\alpha + \beta(1-\alpha) + (1-\beta)(1-\alpha)$, is one.

It was demonstrated in Ref. \citeonline{Jayles2022b} that these inhomogeneities increase recovery time and therefore decrease resilience by our definition of resilience. Therefore, when we fit the two-community model to the survey data, we have to remember that the closer that $\alpha$ and $\beta$ are to 1/2, the more resilient the Singapore society is.

\section{The Survey}
\label{sect:survey}

This survey was approved by the Nanyang Technological University's Institutional Review Board for the period 1 Jun 2022 to 1 Jun 2023, under the reference number IRB-2021-1004.
The survey consisted of two sections, with 13 questions each. The first section asked for demographics information, while the second section comprised questions on smartphone contacts. The full survey questionnaire is included in \ref{app:surveyquestions}. Due to the impending departure of a member (BJ, the main architect of the survey) of our team, we started the survey as soon as we received ethics approval and negotiations with Qualtrics (\url{https://www.qualtrics.com/sg/}) were completed. There were no other reasons for carrying out the survey between 28 Jul 2022 and 15 Sep 2022, with a total of $N = 2,057$ participants identified through representative sampling. Their distributions based on gender, age, and ethnicity are shown in Table~\ref{tab:distributions}.

\begin{table}[htbp]
\tbl{Gender, age, and race distributions of survey participants. For a given criterion, its Quota is the proportion of participants satisfying the criterion, the Target is the maximum number of participants satisfying the criterion (and simultaneously all other criteria), while the Count is the actual number of participants satisfying the criterion. To achieve representative sampling, the Quota must be equal to the proportion of Singapore residents and non-residents satisfying the given criterion.}{
\begin{tabular}{cccc}
\hline
Gender & Quota & Target & Count \\
\hline
Male & 48\% & 1008 & 1008 \\
Female & 52\% & 1092 & 1040 \\
Unknown & 5\% & 105 & 4 \\
\hline
Age & Quota & Target & Count \\
\hline
21-24 & 8\% & 189 & 189 \\
25-34 & 24\% & 546 & 546 \\
35-44 & 23\% & 525 & 497 \\
45-54 & 17\% & 378 & 378 \\
55-64 & 14\% & 315 & 304 \\
65+ & 14\% & 144 & 147 \\
\hline
Race & Quota & Target & Count \\
\hline
Chinese & 75\% & 1575 & 1567 \\
Malay & 13\% & 273 & 249 \\
Indian & 7\% & 147 & 137 \\
Others & 5\% & 105 & 105 \\
\hline
\end{tabular}}\label{tab:distributions}
\end{table}

\begin{figure}[htbp]
\centering
\includegraphics[width=\linewidth]{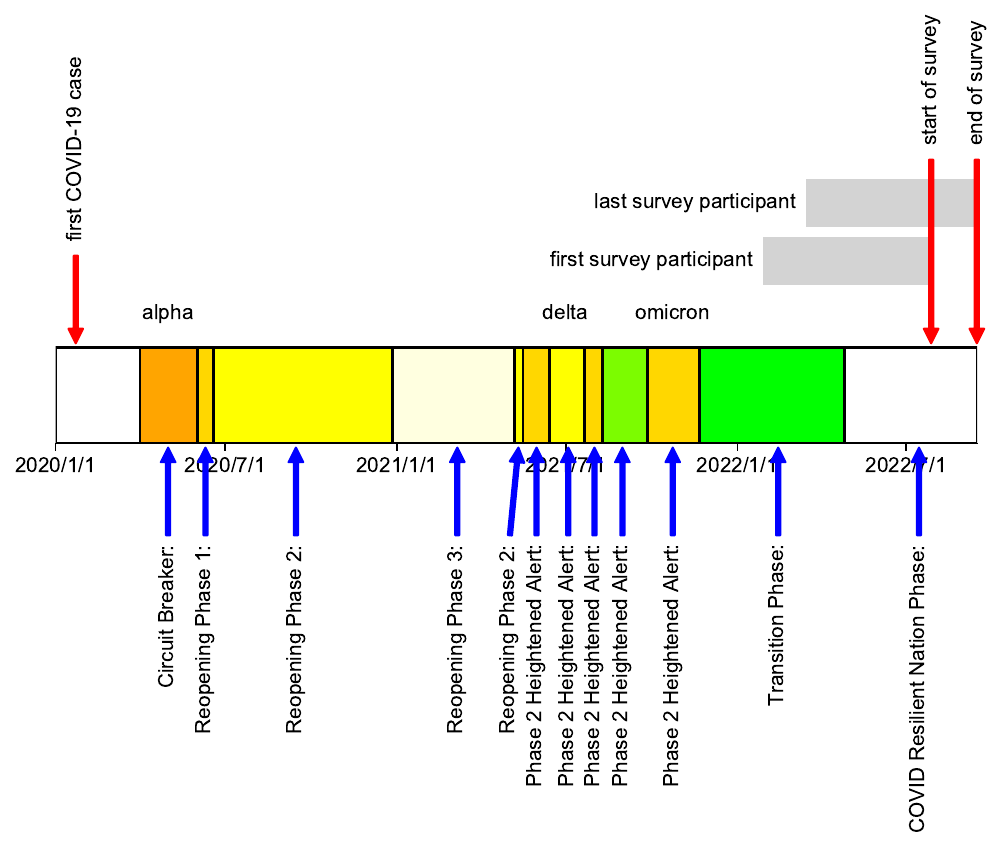}
\caption{Timeline of intervention measures implemented by the Singapore Government during the COVID-19 pandemic and the relative timing of the survey performed for this study. These were introduced in phases with varying levels of restrictions: (1) Circuit Breaker (1 Apr-1 Jun 2020); (2) Reopening Phase 1 (2-18 Jun 2020); (3) Reopening Phase 2 (19 Jun-27 Dec 2020); (4) Reopening Phase 3 (28 Dec 2020-7 May 2021); (5) Reopening Phase 2 (8-16 May 2021); (6) Phase 2 Heightened Alert (16 May-13 Jun 2021); (7) Phase 3 Heightened Alert (14 Jun-21 Jul 2021); (8) Phase 2 Heightened Alert (22 Jul-9 Aug 2021); (10) Preparatory Stage of Transition (10 Aug-26 Sep 2021); (11) Phase 2 Heightened Alert (27 Sep-21 Nov 2021); (12) Transition Phase (22 Nov 2021-25 Apr 2022). From 26 Apr 2022 onwards, Singapore was declared a COVID-19 resilient nation, because it was believed that herd immunity has been achieved by the nation-wide vaccination program. Also indicated in this figure is the first COVID-19 case on 23 Jan 2020, the approximate pandemic periods caused by the alpha, delta, and omicron strains, as well as the start and end of our survey. Finally, we also show in gray the six-month period that would contribute to the response of the first survey participant, and the six-month period that would contribute to the response of the last survey participant.}
\label{fig:COVIDtimeline}	
\end{figure}

In this section, we describe and interpret only the survey questions that were directly used to calibrate our network model. These questions refer to a six-month time period up till the time the participant took the survey. Therefore, the earliest six-month time period in the survey responses is the time interval (28 Jan 2022, 28 Jul 2022), while the latest six-month time period is the time interval (15 Mar 2022, 15 Sep 2022). In Figure \ref{fig:COVIDtimeline} we show when the survey was completed relative to the timeline of the COVID-19 pandemic. Note that the pandemic lockdown in Singapore was lifted shortly before the survey time period, and we argue in Section \ref{sect:discussion} that our simulation indeed shows that the survey results are consistent with an ongoing recovery during the survey. On hindsight, we realized that in line with the flow diagram shown in Figure \ref{fig:SurveyFlow}, it might be better to follow a representative cohort of participants over time, and have them answer questions at the start of the survey, then at the three-month midpoint of the survey, and finally at the six-month endpoint of the survey. However, such a survey would be prohibitively expensive, and beyond what we can afford with our research funds. 

\subsection{One-Community Survey Questions}
\label{section:1commsurv}

To calibrate the one-community model, we relied on five of the 13 smart phone usage questions. These crucial five questions are listed in Table \ref{table:questionsandinterpretations} along with their interpretation as node properties. We will refer to the responses (which are integers) to these five questions as \emph{counts} QB4a, QB4b, QB5, QB6, and QB9.
 
\begin{table}[htbp]
\tbl{Summary of the key survey questions and their interpretations for the purpose of calibrating our models.}{
\begin{tabular}{c p{4.5cm} p{5.5cm}}
\hline
Counts & Question & Interpretation \\
\hline
QB4a & How many different people did you exchange (receive and/or send) at least one message with over the past 6 months? & This is the total number of relationships (links) that existed for the previous 6 months, including links that are removed before the end of the 6 months. \\
\hline
QB4b & And over the past 3 months? & This is the total number of relationships (links) that existed for the most recent 3 months, including links that are removed before the end of the 3 months. \\
\hline
QB5 & Among those people with whom you have exchanged at least one message over the past 6 months, how many had you never exchanged messages with before? & This is the total number of new links acquired during the prior 6 months, including links that were removed before the 6 month period ends. \\
\hline
QB6 & Among people with whom you have exchanged at least one message over the past 6 months, how many did you get to know through other contacts of yours? & This is the total number of links acquired socially during the prior 6 months covered by the survey, combined with any socially-acquired links that survived from before the 6 month period covered by the survey. \\
\hline
QB9 & How many people do you actively communicate with using messaging apps? & This is an effective measure of $z$, the degree of the node, evaluated at the end of the 6 months. \\
\hline
\end{tabular}}\label{table:questionsandinterpretations}
\end{table}

\begin{figure}
    \centering
    \includegraphics[width=1\linewidth]{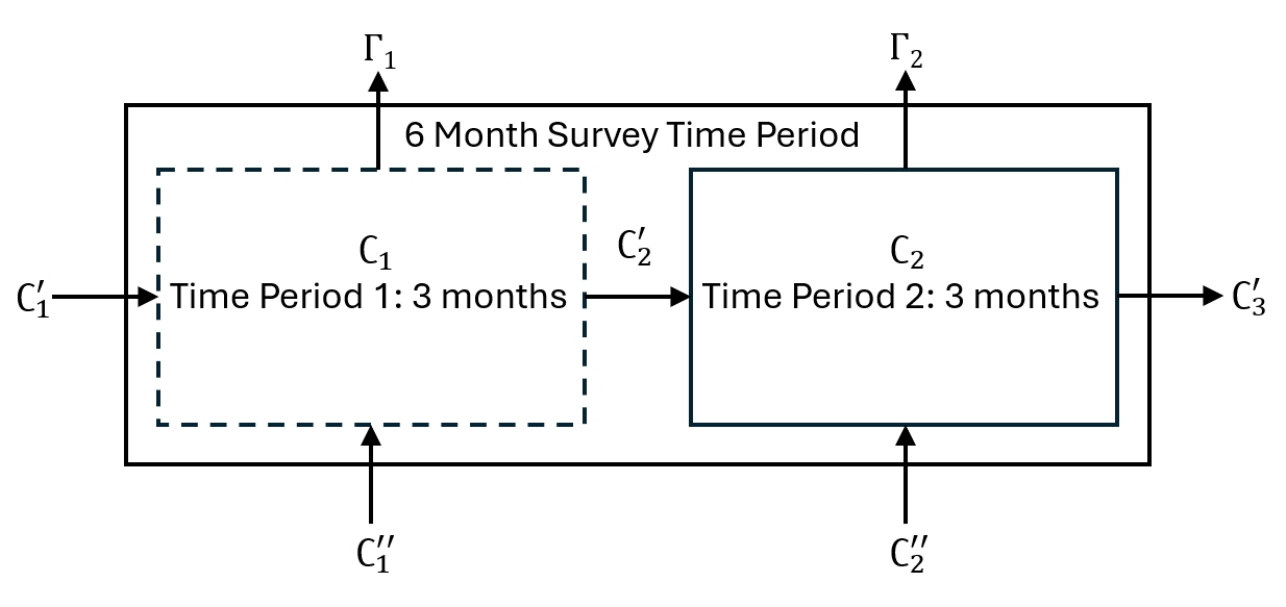}
    \caption{Flow diagram illustrating the time evolution of phone contacts during the six-month time period covered by the survey. A variable without a prime represents the total number of contacts; a variable with a single prime represents the number of contacts that survived from the prior time period; and a variable with a double prime represents contacts that were newly acquired, either randomly or socially during the time period. $\Gamma_1$ represents the number of contacts lost during the first three-month time period, while $\Gamma_2$ represents the number of contacts lost during the second three-month time period.}
    \label{fig:SurveyFlow}
\end{figure}

Our central task is to convert these survey responses into node-level counts of links that existed or were removed during the time period covered by the survey. This task is easier if we follow the flow diagram shown in Figure~\ref{fig:SurveyFlow}. We start the analysis by dividing the six-month interval covered by the survey into two separate time periods, labeled 1 and 2 in Figure \ref{fig:SurveyFlow}. Time Period 2 is represented with a solid-line rectangle because survey question QB4b measures it directly in the survey. On the other hand, Time Period 1 is not measured directly in the survey, and all quantities related to this time period are derived from other quantities. For this reason, it is represented by a dashed rectangle. Lastly, the six-month time period covered by question QB4a is represented by a solid box because it is explicitly studied in the survey. For participant $1 \leq i \leq 2057$, QB4a$_i$, QB4b$_i$, QB5$_i$, QB6$_i$, and QB9$_i$ are his/her responses to survey questions 4a, 4b, 5, 6, and 9 in Section B. Note that in the following analysis, we define extra variables that may seem unnecessary at first, but proved crucial to solving the sets of equations for the quantities of interest.

For simplicity, let us start with Questions B5 and B6. Question B5 asks for the total number of new contacts acquired during the entire six-month period. We can also call this value $R_i = R_{0,i} + R_{1,i}$, where $R_{0,i}$ is the total number of new contacts acquired randomly by participant $i$ during the six-month time period while $R_{1,i}$ is the total number of new contacts acquired socially (through a mutual acquaintance) by participant $i$ during the 6-month time period. Here, let us differentiate between the quantities $R_i$, $R_{0,i}$, $R_{1,i}$, which are \emph{intensive}, because they are associated with each participant, in contrast with the quantities $\mathcal{R}_0$ and $\mathcal{R}_1$ (see Figure \ref{fig:paramcounts}), which are \emph{extensive}, because they include contributions from all nodes (and therefore do not carry the index $i$). From this point on, we will typeset extensive quantities in calligraphic font, and intensive quantities in ordinary italic font, $R$, $R_0$, and $R_1$, suppressing the index $i$ associated with the participants.

In terms of the variables defined in Figure~\ref{fig:SurveyFlow}, we can also write $R = C_1'' + C_2''$, where $C_1''$ is the total number of contacts acquired during Time Period 1, and $C_2''$ is the total number of contacts acquired during Time Period 2. We can therefore summarize the relationships between these variables as
\begin{equation}\label{eq1}
    \text{QB5} = R = R_0 + R_1 = C_1'' + C_2''.
\end{equation}

Next, Question B6 asks for the total number of contacts used during the six-month time period that were acquired from an existing contact. Here, let us note that Question B6 is not asking for contacts that were acquired \emph{during} the six-month time period, so we must assume that many of these contacts were acquired \emph{before} the six-month time period as well as \emph{during} the six-month time period. We can relate this to the variables in Figure \ref{fig:SurveyFlow} by assuming that $C_1'$, the number of contacts acquired before the six-month time period that survive into Time Period 1, is composed of a fraction $f = R_1/R$ of contacts acquired socially.  This analysis can be summarized with the equation
\begin{equation}\label{eq2}
    \text{QB6} = f C_1' + R_1 = f(C_1' + R).
\end{equation}

Now, let us consider Question B4a, which asks for the total number of contacts that a participant has sent messages to or received messages from during the entire six-month time period. This includes contacts acquired before the six-month period, denoted by $C_1'$, and new contacts acquired during the two three-month periods. We can also rewrite this for convenience in terms of $R$ and the total number of contacts $C_1$ with whom the participant exchanged messages with during Time Period 1, as
\begin{equation}\label{eq4}
    \text{QB4a} = C_1' + C_1'' + C_2'' = C_1' + R = C_1 + C_2''.
\end{equation}
Question B4b then asked for the total number of people $C_2$ contacted during Time Period 2, including the number of contacts $C_2'$ acquired before, as well as the number of contacts $C_2''$ acquired during Time Period 2. This tells us that
\begin{equation}\label{eq5}
    \text{QB4b} = C_2' + C_2'' = C_2.
\end{equation}

Finally, we will assume that any contacts who had exchanged messages with the participant during Time Period 1 but not during Time Period 2 have been lost (corresponding to deleted links in the JGN model). This neglects the contacts who contact the participant infrequently. Summing the quantities flowing in and out of the box labeled ``Time Period 1" in Figure \ref{fig:SurveyFlow}, we obtain
\begin{equation}\label{eq5}
     C_2' = C_1' + C_1'' - \Gamma_1 = C_1 - \Gamma_1,
\end{equation}
where $\Gamma_1$ is the total number of contacts ``lost" during Time Period 1 for a particular participant. We have a similar relation for Time Period 2,
\begin{equation}\label{eq5b}
     C_3' = C_2' + C_2'' - \Gamma_2 = C_2 - \Gamma_2,
\end{equation}
wherein $\text{QB9} = C_3'$.

Now that we have this set of five equations, we are prepared to solve for the five unknowns of interest: $\Gamma_1$, $\Gamma_2$, $R_0$, $R_1$, and $C'_3$. These are respectively the total counts of contacts lost in Time Period 1 and Time Period 2, gained randomly, gained socially, or unchanged, for a given participant over the six months covered by the survey. Solving these four parameters in terms of the empirical counts, we first find
\begin{equation}\label{eq6}
\Gamma_1 = \text{QB4a} - \text{QB4b},
\end{equation}
\begin{equation}\label{eq6b}
\Gamma_2 = \text{QB4b} - \text{QB9}.
\end{equation}

Thereafter, using the definition of $f$ being the fraction of contacts acquired socially, we find
\begin{equation}\label{eq7}
    f = \frac{\text{QB6}}{\text{QB4a}},
\end{equation} 
and thus
\begin{equation}\label{eq8}
    R_1 = f R = \frac{\text{QB6}\cdot\text{QB5}}{\text{QB4a}}.
\end{equation}
Finally, we find
\begin{equation}\label{eq9}
    R_0 = \text{QB5}\left(1 - \frac{\text{QB6}}{\text{QB4a}}\right).
\end{equation}

We also define
\begin{equation}
    \text{CORE} = \text{QB4a} - \text{QB5} - \Gamma_1 - \Gamma_2,
\end{equation}
which represents the number of contacts acquired before the six-month survey time period and remained unchanged after the six-month survey time period. Some participants will have a negative value of $\text{CORE}$, some will have a positive value, and some will have a value of zero. The median is 1 and the mean is about 3, likely accounting for close family connections that remain unchanged.

Using the \texttt{Dataframes} package in Python, and the method \texttt{dataframe.describe()}, we obtained the summary statistics shown in Table \ref{table:VertexProps}. As we can see from these basic statistical measures of the responses and our derived quantities, the mean differs significantly from the median and also has a very large standard deviation. Both of these features are due to very fat tails in the distributions. For this reason, we chose to work with the median in all of our measurements. Later we will also find it useful to compute the total contacts during Time Period 1,
\begin{equation}\label{eq10}
    C_1 = \text{QB4a} - \frac{\text{QB5}\cdot\text{QB4b}}{\text{QB4a}}.
\end{equation}
Note that $C_2$ is already determined by Question QB4b.

\begin{table}[htbp]
\tbl{Summary of model-relevant survey measurements, displaying mean, median, and several measures of distribution about the averages. Columns with a count less than 2,057 had some missing values. The standard deviation reported in this table is that of the sample of 2,057 participants. It is also reasonable to use the standard deviation $\text{std}/\sqrt{\text{count}}$ of the sample mean as an estimate of the error in the mean value. Although some intensive values are negative, the averages are all positive and consistent with our model and also with common sense.}{
\begin{tabular}{ccccccccccccc}
\hline
Index & QB4a & QB4b & QB5 & QB6 & QB9 & $\Gamma_1$& $\Gamma_2$ & $f$ & $R_0$ & $R_1$  & $\text{CORE}$\\
\hline
count & 2057 & 2057 & 2047 & 2047 &  2057   & 2057 &   2057    & 2047 & 2047 & 2047  & 2057\\
mean & 38.17 & 28.72 & 16.48 & 13.15  &   19.38   & 9.45 & 9.35  & 0.40 & 8.57 & 7.90  & 2.98\\
std & 87.07 & 66.17 & 50.71 & 47.91 &  47.35   & 33.03 &   50.82  & 0.37 & 30.94 & 34.95  & 50.27 \\
min & 0 & 0 & 0 & 0 & 0   &  0   &   $-250$  & 0.00 & 0 & 0.0  & -995 \\
25\% & 7 & 5 & 2 & 1 &  4   & 0 &   0  &  0.10 & 0 & 0.1  & -1\\
50\% & 20 & 12 & 5 & 4 &  10   & 1 &  1   & 0.27 & 2 & 1.25  & 1\\
75\% & 40 & 30 & 15 & 10 &  20  &  10 &  10   & 0.70 & 7.5 & 5.0  & 8\\
max & 1000 & 1000 & 1000 & 1000 &   1000   & 750 &   995   & 1.00 & 665 & 1000  & 500\\
\hline
\end{tabular}}\label{table:VertexProps}
\end{table}

\subsection{Two-Community Survey Questions}
\label{section:2commsurvey}

As described in Section \ref{sect:twoJGN}, we can build a two-community model by dividing the population into two separate communities, distinguished by two additional model parameters $\alpha$ and $\beta$. In this subsection, we focus on the portion of the survey relevant to evaluating these inter-community and intra-community parameters. There are seven possible demographic categories we can use to do this. However, to provide a single concrete example throughout the rest of this paper, we will focus on age, i.e., a community of older people (denoted by $B$) coexisting with a community of younger people (denoted by $A$), because of the extensive literature suggesting generational differences in mobile apps\cite{zhitomirsky2016cross,hur2017understanding} and social media\cite{leung2013generational,vittadini2013generations,fietkiewicz2016inter} usage. 

\begin{figure}
    \centering
    \includegraphics[width=0.9\linewidth]{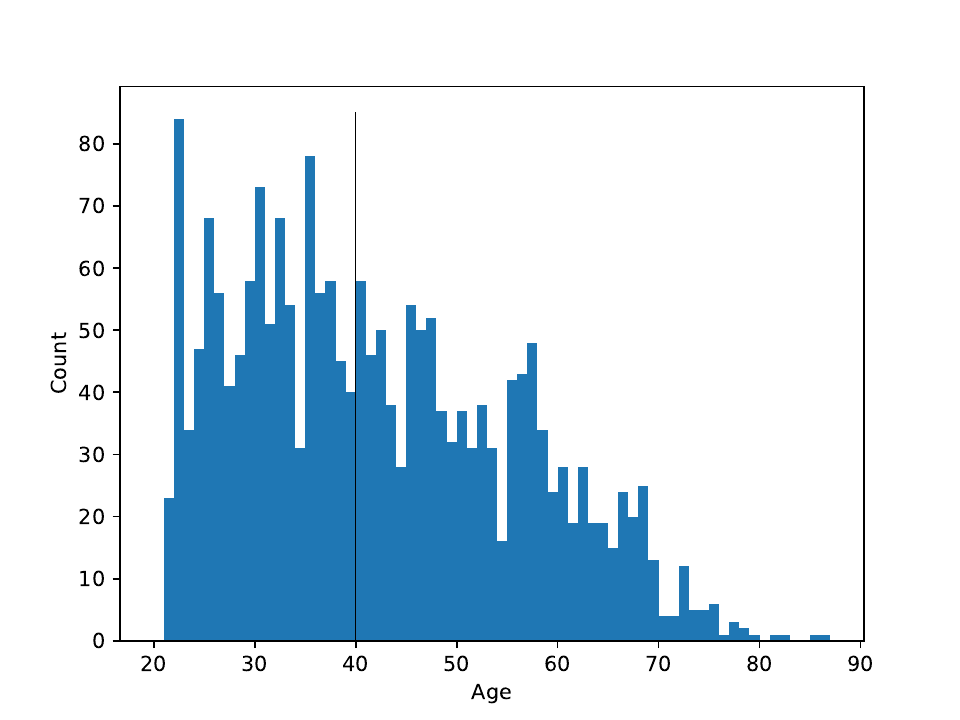}
    \caption{The complete distribution of age from Question A2 of our survey. The Singapore Department of Statistics maintains a database of the number of residents by age, from 0 years of age to 89 years of age, and the histogram of this demographics data for 2023 is smoother than what we show here. This is because for statistical analyses, the Singapore Department of Statistics recommends the use of five-year age groups as part of their National Statistical Standards. As we can see in Table \ref{tab:distributions}, Qualtrics selects survey participants based on these age groups, and therefore they do not control the proportions down to the actual age. The Qualtrics survey also does not include participants below 21 years of age, because the standards of ethics approval to include minors are much higher.}
    \label{fig:histAge}
\end{figure}

First, let us focus on the intra-community parameter $\beta$ by splitting the population into two roughly equal-sized communities, using the median age of 40 from Question A2 in the survey to break the $N = 2,057$ participants into one community strictly below the median age, and another community equal to or above the median age. In Figure \ref{fig:histAge}, we show the age distribution of our survey participants, and a black vertical line showing the median age of 40. Once we split the dataset into two communities (in our case, using a simple \texttt{Dataframes} command to create a separate \texttt{Dataframe} for each community), we run the one-community analysis described in Section \ref{sect:onecommunitycalibration} for each of the communities to obtain $R_0$, $R_1$, and $\Gamma$ for each of the communities. 
Thereafter, we use the $R_0$'s obtained for each community to estimate the parameter $\beta$.

\begin{figure}[htbp]
    \centering
    \includegraphics[width=0.7\linewidth]{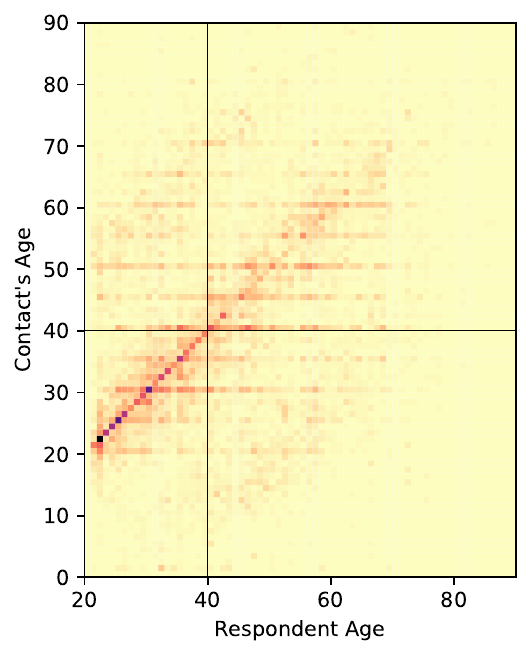}
    \caption{Heat map of the distribution of the contact's ages obtained from Question B11a. In this figure, the horizontal axis is the participant's age (from Question A2), while the vertical axis is the contact's age (from Question B11a). In this plot, we see strong evidence for a participant is more likely to be connected to someone close to his/her own age, to his/her parents 20-30 years older, to his/her children 20-30 years younger.}
    \label{fig:histAge2D}
\end{figure}

To compute the inter-community parameter $\alpha$ we will need to correlate the two communities. This analysis is enabled by Question B11a. Question B11 asks: ``Think about the 5 persons with whom you communicate most regularly:" and part (a) asks: ``How old are they?" with five spaces. The results of this question are summarized in Figure \ref{fig:histAge2D}, where we plot the distribution $f(m,n)$ of the number of participants with age $m$ (horizontal axis) with contact of age $n$ (vertical axis). 

After aggregating the histogram in Figure \ref{fig:histAge2D} based on the division of participants into communities $A$ and $B$, we can then estimate the probabilities $P(A, A)$, $P(A, B)$, and $P(B, B)$ for links in the two different communities. To estimate these probabilities, we assume that at steady state, a member $i$ in community $A$ should have $P(A, B) z_i$ contacts in community $B$, if he/she has $z_i$ contacts in total. Furthermore, because the rate of link creation is controlled by $r_0$, we can say that a member of community $A$ acquires links to members of community $B$ at a rate of $P(A, B) r_0$. Comparing this with the description of the two-community model in Section \ref{sect:twoJGN}, we see that $P(A, B) = \alpha$. For participant $i$ in community $A$, we compute $P_i(A,B) = z_{i,B}/\tilde{z}_i$, where $z_{i,B}$ is the number of contacts from community $B$. Here, we use $\tilde{z}_i = 5$ instead of $z_i$ from Question B9, because this is the number of possible responses in Question B11a. After computing this value for all members of community $A$, we take the average to be $P(A, B)$. 

\section{Calibrating the One- and Two-Community Models}
\label{section:methods}

In this section, we explain how we calibrate the one- and two-community models using the survey data. This is a multi-step process. We will start by describing in Section \ref{sect:onecommunitycalibration} our custom statistical analysis of the one-community model data, before describing in Section \ref{sect:simulationmapping} the simulation-based procedure to match model parameters to the survey analysis outputs. We complete the calibration of the two-community model in Section \ref{sect:twocommunitycalibration}.

\subsection{Estimating the Model Counts}
\label{sect:onecommunitycalibration}

In Section \ref{section:1commsurv}, we solved for the model counts $R_0$, $R_1$, $\Gamma_1$ and $\Gamma_2$ in terms of counts QB4a, QB4b, QB5, QB6, and QB9 obtained from the survey. We summarize them here as
\begin{align*}
\Gamma_1 &= \text{QB4a} - \text{QB4b}, \\
\Gamma_2 &= \text{QB4b} - \text{QB9}, \\
R_0 &= \text{QB5}\left(1 - \frac{\text{QB6}}{\text{QB4a}}\right), \\
R_1 &= \frac{\text{QB6}\cdot\text{QB5}}{\text{QB4a}}.
\end{align*}
From 10,000 bootstraps (see the textbook by Efron and Tibshirani\cite{Efron1993} on the bootstrap method, and the SciPy online user's manual\cite{ScipyBootstrap} on how this can be done in Python.), we estimated the medians of these model counts and their standard errors, as shown in Table \ref{table:estimates}. For completeness sake, we also included the survey counts, QB4a, QB4b, QB5, QB6, QB9, and the counts $C_1$, $C_2$, $C'_1$, $C'_2$, $C''_1$, $C''_2$ shown in Figure \ref{fig:SurveyFlow}. We checked that these estimates do not change when we increase the number of bootstraps to 100,000.

\begin{table}[htbp]
\tbl{Summary of model counts. In the last column, the count is less than 2,057 if no value was reported in Question B5 or Question B6.}{
\begin{tabular}{cccccc}
\hline
 & median & error & 95\% low & 95\% high & count \\
\hline
QB4a & 20.0 & 1.4 & 15.0 & 20.0 & 2057 \\
QB4b & 12.0 & 1.3 & 10.0 & 15.0 & 2057 \\
QB5 & 5.0 & 0.0 & 5.0 & 5.0 & 2057 \\
QB6 & 4.00 & 0.41 & 3.0 & 4.0 & 2057 \\
QB9 & 10.0 & 0.61 & 8.0 & 10.0 & 2057 \\
$R_0$ & 2.00 & 0.21 & 2.0 & 2.5 & 2047 \\
$R_1$ & 1.25 & 0.11 & 1.0 & 1.5 & 2047 \\
$\Gamma_1$ & 1.0 & 0.2 & 1.0 & 1.0 & 2057 \\
$\Gamma_2$ & 1.0 & 0.4 & 1.0 & 2.0 & 2047 \\
$f$ & 0.27 & 0.02 & 0.25 & 0.3 & 2047 \\
$C_1$ & 10.00 & 0.13 & 9.5 & 10.0 & 2047 \\
$C'_1$ & 8.0 & 0.6 & 7.0 & 9.0 & 2057 \\
$C''_1$ & 0.0 & 0.0 & 0.0 & 0.0 & 2047 \\
$C_2$ & 12.00 & 1.27 & 10.0 & 15.0 & 2057 \\
$C'_2$ & 5.83 & 0.42 & 5.0 & 6.4 & 2047 \\
$C''_2$ & 4.0 & 0.1 & 3.75 & 4.15 & 2047 \\
$R_0 + R_1 + \Gamma_1$ + $\Gamma_2$ & 12.0 & 0.93 & 10.0 & 13.0 & 2047 \\
$z$ & 7.00 & 0.42 & 6.3 & 8.0 & 2047 \\
$\text{CORE}$  & $1.0$ & $0.44$ & $1.0$ &  $2.0$ &  $2057$ \\
\hline
\end{tabular}}\label{table:estimates}
\end{table}

\subsection{Simulation Matching the One-Community Model Parameters}
\label{sect:simulationmapping}

\begin{figure}[htbp]
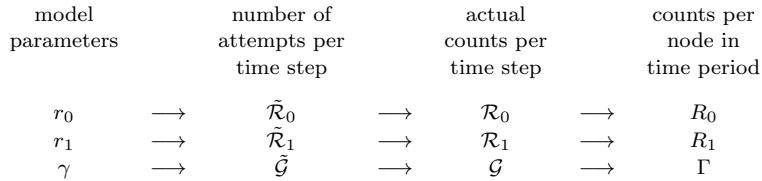

\centering\footnotesize
\begin{tabular}{ccccccc}
model & & number of & & actual & & counts per \\ 
parameters & & attempts per & & counts per & & node in \\
 & & time step & & time step & & time period \\ [2ex]
$r_0$ & $\longrightarrow$ & $\tilde{\mathcal{R}}_0$ & $\longrightarrow$ & $\mathcal{R}_0$ & $\longrightarrow$ & $R_0$ \\
$r_1$ & $\longrightarrow$ & $\tilde{\mathcal{R}}_1$ & $\longrightarrow$ & $\mathcal{R}_1$ & $\longrightarrow$ & $R_1$ \\
$\gamma$ & $\longrightarrow$ & $\tilde{\mathcal{G}}$ & $\longrightarrow$ & $\mathcal{G}$ & $\longrightarrow$ & $\Gamma$\\ [1ex]
\end{tabular}
\caption{Relationship between the mJGN model parameters $r_0$, $r_1$, $\gamma$, the extensive number of attempts $\tilde{\mathcal{R}}_0$, $\tilde{\mathcal{R}}_1$, $\tilde{\mathcal{G}}$ sampled from Poisson distributions with means $r_0 N$, $r_1 N_m$, $\gamma N_e$, the extensive actual counts $\mathcal{R}_0$, $\mathcal{R}_1$, $\mathcal{G}$ and intensive counts per node $R_0$, $R_1$, and $\Gamma$. The first two sets of counts are over all nodes, for a single time step, whereas the last set of counts are over a time period, for a single node.}
\label{fig:paramcounts}
\end{figure}

Up to this point, we have estimated the average model counts $R_0$, $R_1$, $\Gamma_1$, $\Gamma_2$ per participant in the survey time period, but what we would like to estimate are the model parameters $r_0$, $r_1$, $\gamma$. As shown in Figure \ref{fig:paramcounts}, the extensive numbers of attempts $\tilde{\mathcal{R}}_0$, $\tilde{\mathcal{R}}_1$, $\tilde{\mathcal{G}}$ are sampled each time step from Poisson distributions with means $r_0 N$, $r_1 N_m$, and $\gamma N_e$ respectively. If we have the distributions of $\tilde{\mathcal{R}}_0$, $\tilde{\mathcal{R}}_1$, and $\tilde{\mathcal{G}}$, we could estimate $r_0$, $r_1$, and $\gamma$ from the means of these distributions.

However, not all attempts are successful. The actual numbers of counts $\mathcal{R}_0$, $\mathcal{R}_1$, $\mathcal{G}$ per time step are also extensive, but they are no longer directly related to the parameters $r_0$, $r_1$, $\gamma$. In general, $\mathcal{R}_0 < \tilde{\mathcal{R}}_0$ and $\mathcal{R}_1 < \tilde{\mathcal{R}}_1$, because sometimes the attempt to create a link for nodes $i$ and $j$ fails as $i$ and $j$ are already linked, or $z_i = z^*$, or $z_j = z^*$, and thus no new link can be added. Analogously, $\mathcal{G} \leq \tilde{\mathcal{G}}$ because the same nodes can be selected repeatedly for link deletion. If the link from node $i$ is deleted, subsequent attempts to delete the same link will fail.

Ideally, our survey with $N = 2,057$ participants samples a social network (i.e., the Singapore population) of 5.5 million nodes. If we believe that our survey is representative, our rate of random links added per node $\braket{R_0}$ and rate of social links added per node $\braket{R_1}$ are intensive quantities, and should be the same whether obtained by averaging over 2,057 participants, or over 5.5 million people. In Refs. \citeonline{Jayles2022a} and \citeonline{Jayles2022b}, $\braket{\mathcal{R}_0}$ and $\braket{\mathcal{R}_1}$ represent the total rate of random links added per time step, and since they were found to both scale as $N$, $\braket{R_0} = \left(T/N\Delta t\right)\braket{\mathcal{R}_0}$ and $\braket{R_1} = \left(T/N\Delta t\right)\braket{\mathcal{R}_1}$ are indeed intensive quantities. Here, $N$ is the number of nodes, $\Delta T$ is the simulation time step, and $T$ is the duration of the measurement. We did not check whether $\braket{\mathcal{G}}$ scales with $N$ in Refs. \citeonline{Jayles2022a} and \citeonline{Jayles2022b}, but it is reasonable to assume that $\braket{\Gamma} = \left(T/N\Delta t\right)\braket{\mathcal{G}}$ are also intensive.

This suggests that to estimate $r_0$, $r_1$, and $\gamma$, we should run an ensemble of mJGN simulations of many nodes for different combinations of $r_0$, $r_1$, and $\gamma$, and choose the combination that most closely reproduces $\braket{R_0} = 2.00 \pm 0.21$, $\braket{R_1} = 1.25 \pm 0.11$, and $\braket{\Gamma} = \braket{\Gamma_1} + \braket{\Gamma_2} = 2.0 \pm 0.6$. In fact, we expect the same combination of $r_0$, $r_1$, $\gamma$ would produce similar values of $\braket{R_0}$, $\braket{R_1}$, $\braket{\Gamma}$, whether we run the simulations on a $N = 50$ network or on a $N=5,500,000$ network.

Therefore, for our simulation-based calibration, we ran 100 mJGN simulations of 1,000 time steps with $(N = 50, z^* = 5)$ for each parameter combination in a step-wise logarithmic grid of points in the $(r_0, r_1, \gamma)$ parameter space, with 18 different values for $r_0$, 20 values for $r_1$, and 20 values for $\gamma$, totaling about $8,000$ parameter combinations. 1,000 time steps was long enough for the simulation to reach steady state, and the complete set of simulations took one week on a single laptop. After a simulation reached steady state, we measure the extensive quantities $\mathcal{R}_0$, $\mathcal{R}_1$, $\mathcal{G}$, and the average clustering coefficient
\begin{equation}
C = \frac{1}{N}\sum_i \frac{e_i}{\frac{1}{2} z_i (z_i - 1)},
\end{equation}
where $e_i$ is the number of edges between neighbors of node $i$. For each parameter combination, we average these quantities over the 100 simulations.

Since optimizing three parameters simultaneously is technically very demanding, we decided to run the calibration to match only $\braket{R_0} = 2.00 \pm 0.21$ and $\braket{R_1} = 1.25 \pm 0.11$, accepting the slight risk that we will over-estimate $r_0$ and $r_1$, while under-estimating $\gamma$. For the parameter combination $(r_0, r_1, \gamma)$, we measure the average counts $\braket{\mathcal{R}_0}$ and $\braket{\mathcal{R}_1}$ per time step over $N = 50$ nodes. To decide whether this is a good fit, we divide $\mathcal{R}_0$ and $\mathcal{R}_1$ by $N$, and then multiply them by the number of time steps $T/\Delta T$. While the period $T$ is known to be 182 days, we cannot uniquely specify the simulation time step $\Delta T$: a simulation with $(r_0, r_1, \gamma)$ over $T$ time steps of size $\Delta T$ is equivalent to a simulation $(r_0/2, r_1/2, \gamma/2)$ over $2T$ time steps of size $\Delta T/2$. Given $\braket{\mathcal{R}_0}/N$ and $\braket{\mathcal{R}_1}/N$, we can determine the number of time steps $T/\Delta T$ required for a match, by dividing $\braket{R_0}$ by $\braket{\mathcal{R}_0}/N$, or $\braket{R_1}$ by $\braket{\mathcal{R}_1}/N$, or $\braket{R_0} + \braket{R_1}$ by $\braket{\mathcal{R}_0}/N + \braket{\mathcal{R}_1}/N$. If we use the last method to determine $T/\Delta T$, we find good fits when the ratio $\braket{R_1}/\braket{R_0}$ is close to $r_1/r_0$.

The best fit is the parameter combination $(r_0 = 0.01, r_1 = 0.002, \gamma = 0.005)$, with average clustering coefficient of $C = 0.142$, and a corresponding time step size of $\Delta T = 0.475$ day.
 
\subsection{Estimating the Two-Community Parameters}
\label{sect:twocommunitycalibration}

Let us assume that $(r_0, r_1, \gamma)$ in the two-community model are the same as in the one-community model, and just estimate the parameters $\alpha$ and $\beta$ unique to the two-community model. From the plot of the responses to Question B11a against the corresponding responses to Question A2 in Figure \ref{fig:histAge2D}, we can extract correlations between the ages of contacts and the ages of survey participants. This allows us to estimate $P(A, B)$, which we assume is equal to $P(B, A)$, and calculate
\begin{equation}\label{eqn:estimatealpha}
    \alpha = \frac{N_{A}P(A, B) + N_{B}P(B, A)}{N},
\end{equation}
where community $A$ consists of all survey participants less than or equal to 39 years of age, while community $B$ consists of all survey participants above the age of 39. Here, $N = 2,057$ is the number of survey participants, $N_A = 1,011$ is the number of young participants, and $N_B = 1,046$ is the number of old participants.

We then obtained the probabilities listed in Table \ref{table:2commageprob} by splitting Figure \ref{fig:histAge2D} into quadrants and sum over entries in each quadrant, so that $P(A, B)$ is the sum of entries in the upper left quadrant divided by the sum of entries in the left half of the figure. Likewise, $P(A, A)$ is the sum of entries in the lower left quadrant divided by the sum of entries in the left half of the figure. Using these values in Eq. \eqref{eqn:estimatealpha} we get $\alpha = 0.305$, indicating a strong preference for intra-community links over inter-community links.

\begin{table}[htbp]
\tbl{Connection probabilities between age groups $A$ and $B$.}{
    \begin{tabular}{|c|c|}
    \hline
    $P(B, A) = 0.316$   & $P(B, B) = 0.683$ \\
    \hline
    $P(A, A) = 0.707$  & $P(A, B) = 0.292$ \\
   \hline
    \end{tabular}}\label{table:2commageprob}
\end{table}

Moving on, let us estimate the intracommunity parameter $\beta$. Here, let us remind readers that $\beta$ is related to the total number of attempted link creations \emph{in the model simulation}, i.e.,
\begin{equation}
    \beta \tilde{\mathcal{R}}_0 + (1-\beta)\tilde{\mathcal{R}}_0 = \tilde{\mathcal{R}}_0,
\end{equation}
which we can rewrite as
\begin{equation}
    \tilde{\mathcal{R}}_A + \tilde{\mathcal{R}}_B = \tilde{\mathcal{R}}_0.
\end{equation}

From our survey, we estimated the intensive quantity $R_0$, which is the average number of random links created per participant. If the participants are not from a single homogeneous community, but from two communities $A$ and $B$ with different average number of random links created per participant,
\begin{equation}\label{eqn:betaA}
    R_A = \beta_A R_0,
\end{equation}
and
\begin{equation}\label{eqn:betaB}
    R_B =  \beta_B R_0,
\end{equation}
then their weighted average would be
\begin{equation}\label{eqn:betaAB}
    \beta = \frac{N_A\beta_A + N_B\beta_B}{N}.
\end{equation}
Using \eqref{eqn:betaA} and \eqref{eqn:betaB}, we can rewrite \eqref{eqn:betaAB} as
\begin{equation}
    \beta = \frac{N_AR_A + N_B(R_0-R_B)}{NR_0}.
\end{equation}

From the survey results, we found $R_A = 2.4$ and $R_B = 2.0$. Having $R_B = R_0$ poses a problem, because $\beta_B$ would be zero. Therefore, we replaced the one-community value of $R_0 = 2.0$ by $R_0 = 2.2$, the average over the young and old communities. With this value of $R_0$, we get $\beta = 0.582$ and $1 - \beta = 0.418$, which tells us that the rate of random link creation is higher in the young community. This agrees with what we expect intuitively. In this sense, we can also think of the intra-community parameter $\beta$ as a \emph{heterogeneity} parameter specifying the behavioral differences between the two communities.

\section{Discussion}
\label{sect:discussion}

\begin{figure}[htbp]
    \centering
    \includegraphics[width=0.9\linewidth]{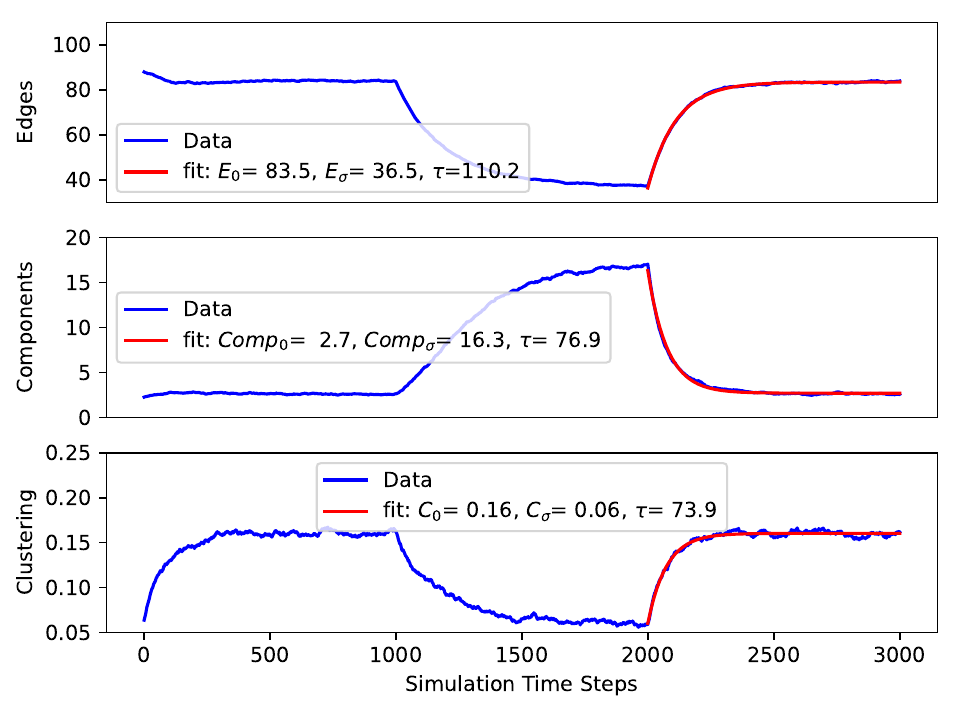}
    \caption{Full lockdown simulation of our model with ($r_0$, $r_1$, $\gamma$) = (0.01, 0.002, 0.005) as calibrated from a survey of Singapore's social network. We can see how the number of links, clustering coefficient, and number of components evolve with time. The edge recovery time is 111 simulation time steps, or approximately 53 human days. Note that all measures of network health are worsened during the lockdown: half the edges are lost; the number of components increases so that the groups consist of just several people; and clustering drastically decreases so that friends of friends do not know each other.}
    \label{fig:lockdownBaseModel}
\end{figure}

In this section, let us check whether the Singapore society is indeed resilient, at least within a one-community approximation. The most convincing way to do this would be to run lockdown simulations showing how the Singapore society bounces back. Starting from random initial networks with $N = 50$, we ran 3,000 simulations with $(r_0 = 0.01, r_1 = 0.002, \gamma = 0.005)$. We ran 1,000 time steps for the simulations to reach steady state, before imposing a 1,000-time-step lockdown (roughly 16 months in the real world), during which the values of $r_0$ and $r_1$ were reduced by a factor of $\sigma = 3$. After the lockdown was lifted, we ran the simulations for another 1,000 time steps for the networks to recover to the steady state. During this recovery, we measured the number of links, number of connected components, and the average clustering coefficient, and fitted the recoveries (averaged over 3,000 simulations) to exponential curves of the form $y(t) = y_0 + (y_{\sigma} - y_0) e^{-t/\tau}$, as shown in Figure \ref{fig:lockdownBaseModel}.

\begin{figure}[htbp]
    \centering
    \includegraphics[width=0.9\linewidth]{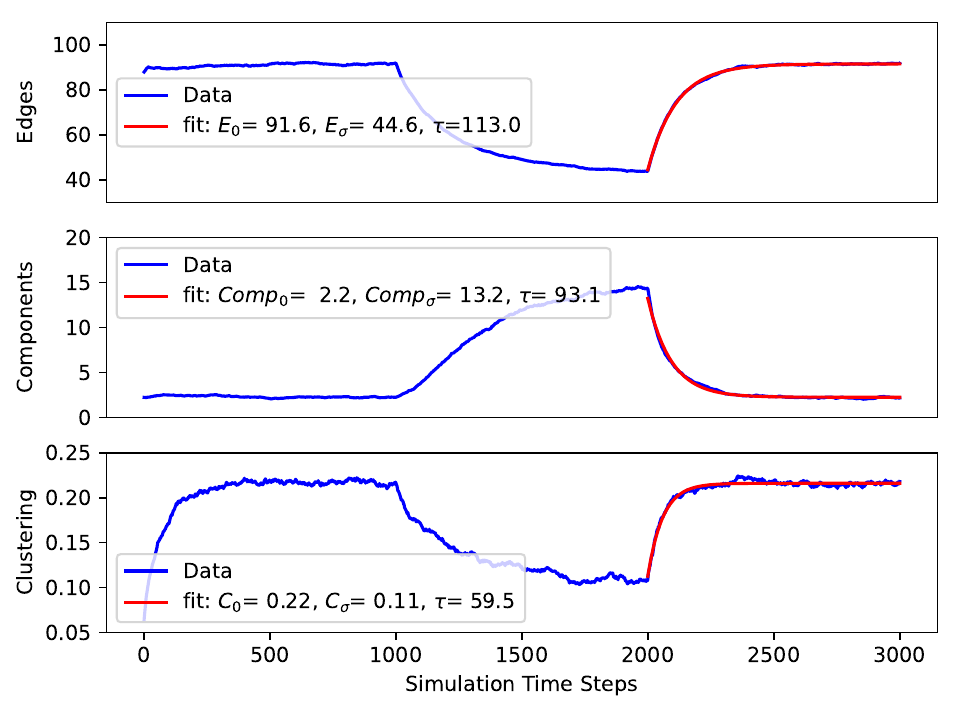}
    \caption{Base model but with modified input parameters, ($r_0$, $r_1$, $\gamma$) = (0.01, 0.004, 0.005). Parameter $r_1$ has been doubled with respect to Figure \ref{fig:lockdownBaseModel} to increase the rate of social links. This leads to an overall increase in the clustering but no improvement in the recovery time and also no change in the resistance to damage during the lockdown. However, the overall clustering is higher both before and during the lockdown.}
    \label{fig:lockdownModified1}
\end{figure}

\begin{figure}[htbp]
    \centering
    \includegraphics[width=0.9\linewidth]{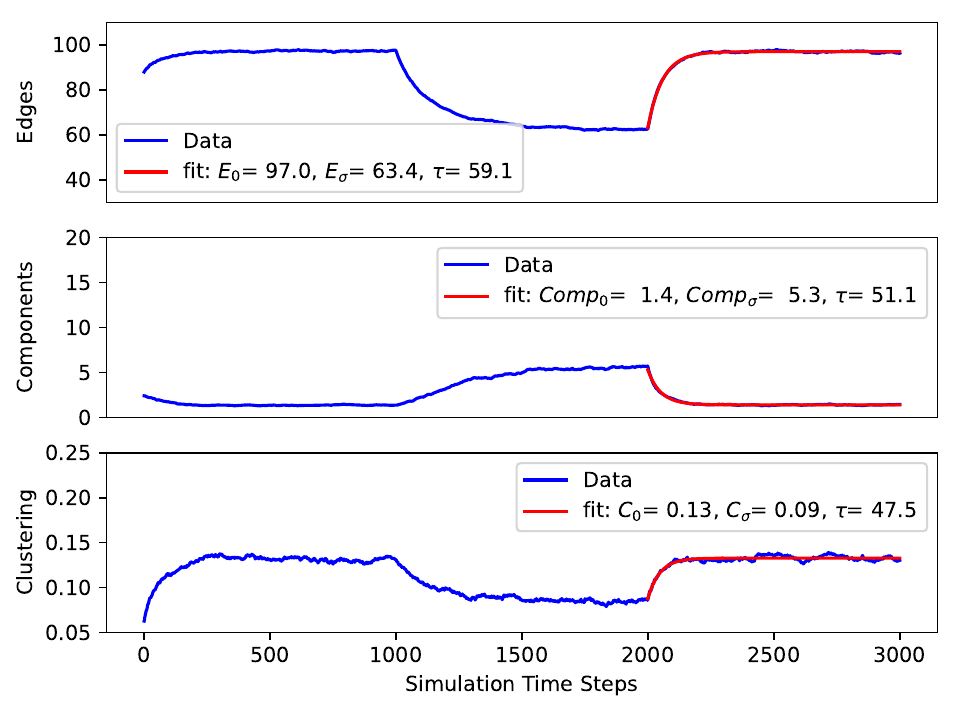}
    \caption{Base model but with modified input parameters, ($r_0$, $r_1$, $\gamma$) = (0.02, 0.002, 0.005), with $r_0$ doubled with respect to Figure \ref{fig:lockdownBaseModel} to increase the rate of random links. This leads to a signficant reduction in recovery time as well as to higher resistance to damage during the lockdown.}
    \label{fig:lockdownModified2}
\end{figure}

To check the relative importance of $r_0$ and $r_1$, we also ran a first set of 3,000 simulations, where $r_0$ and $\gamma$ remained the same, but $r_1$ was doubled from $r_1 = 0.002$ to $r_1 = 0.004$, a second set of 3,000 simulations, where $r_1$ and $\gamma$ remained the same, but $r_0$ was doubled from $r_0 = 0.01$ to $r_0 = 0.02$, and a third set of 3,000 simulations, where both $r_0$ and $r_1$ were doubled. As shown in Figure \ref{fig:lockdownModified1}, when $r_1$ was doubled, only the average clustering coefficient recovered faster. The number of links and the number of connected components both recovered slightly slower. On the other hand, when $r_0$ was doubled, all three quantities recovered faster, as shown in Figure \ref{fig:lockdownModified2}. Finally, when both $r_0$ and $r_1$ were doubled, the recovery rates of the number of links and the number of connected components remained roughly the same as when only $r_0$ was doubled, but the average clustering coefficient recovered dramatically faster. Increasing both parameters leads to the greatest overall improvement in network recovery time, but has the additional benefit of increasing the number of links, decreasing the number of components, and increasing clustering both with and without lockdowns. However, if only one parameter could be targeted for interventions, we recommend $r_0$.

\begin{figure}[htbp]
    \centering
    \includegraphics[width=0.9\linewidth]{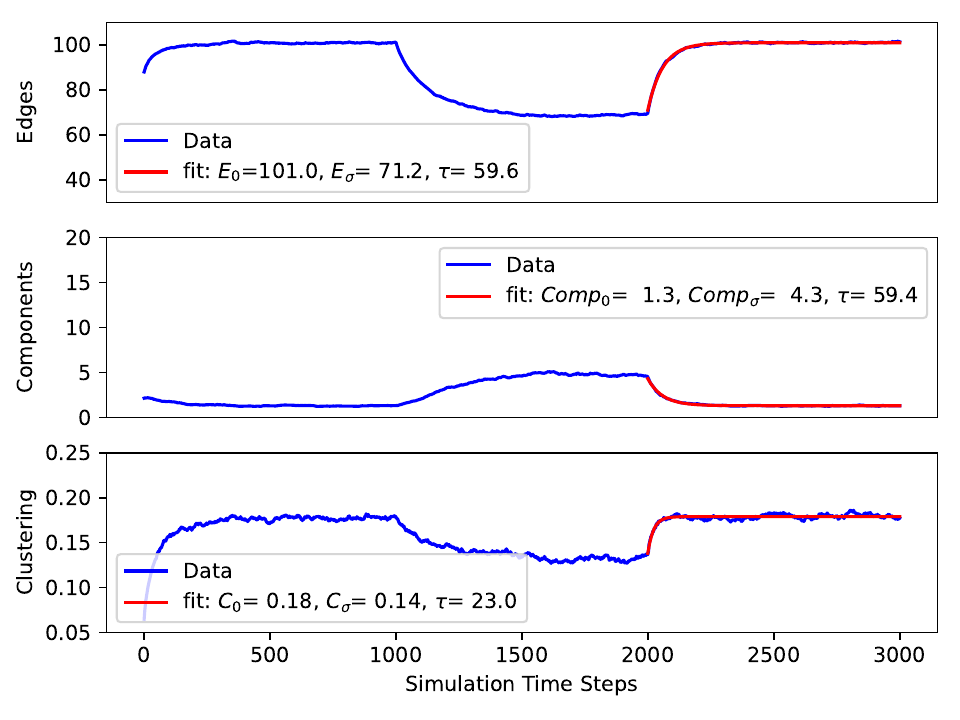}
    \caption{Base model but with modified input parameters, ($r_0$, $r_1$, $\gamma$) = (0.02, 0.004, 0.005), with $r_0$ and $r_1$ doubled with respect to Figure \ref{fig:lockdownBaseModel} to increase the rates of both random and social links. Increasing both parameters leads to improvements in both resistance to damage from the lockdown as well as decreased recovery time as well as increases in all measures of network health: more links, higher clustering, and fewer components, during and before the lockdown.}
    \label{fig:lockdownModified3}
\end{figure}

In these scenario simulations, the number of links reached its pre-lockdown value between 1--2 months after lockdown is lifted. Since we have simulated the best-fit model, we suspect the number of links in the actual Singapore society would also recover within 1--2 months. This sounds fairly resilient to us. We did not repeat this resilience analysis with the best-fit two-community model, since we know from $\alpha = 0.305$ and $\beta = 0.582$ that it would be less resilient than the one-community model.

Let us discuss the three main challenges we encountered in this study. First, there is the issue of \emph{heterogeneity}. In the JGN model, all nodes are equivalent, even though at any given time step, they can have different degrees $z_i$. In principle, we can incorporate the three update rules in the JGN model into a master equation describing the time evolution of the degree distribution $f(z,t)$, and solve for the steady-state distribution $f^*(z)$. Master equations such as this have been extensively studied \cite{leyvraz2003scaling,krapivsky2010kinetic,banasiak2019analytic,villermaux2020fragmentation,singh2022challenges}, and the steady-state distributions can be decaying exponentials or power laws. The empirical degree distribution we obtained from the survey cannot be fitted to any of these, but this is not surprising: unlike in the JGN, where individual nodes have the same rates of gaining random and social connections, or losing connections, in real-world Singapore the individuals probably have broadly different socialities. In fact, as far as the use of social media app is concerned, there seems to be a small population of gregarious individuals who actively maintains a large number of connections. In this paper, we model the Singapore population as a homogeneous social network, or as one containing two homogeneous communities. The two communities we modeled are roughly equal in size, but they are not vastly dissimilar in sociality. At this point, we cannot think of a good way to model the observed heterogeneity, and do not know what effects this heterogeneity might have on the social resilience of Singapore.

The second challenge we encountered is the stationarity of the survey data. The model we fit the data to is statistically stationary, after it is allowed to reach its steady state. This means that $\mathcal{R}_0$, $\mathcal{R}_1$, and $\mathcal{G}$ will have approximately the same values, whenever we perform measurements in the steady state. As we can see from Figure \ref{fig:COVIDtimeline}, the start of the six-month period that survey participants were asked about occurred in the Transition Phase, when some social restrictions were still in place. Even if we accept based on the results of our lock down simulations that the social recovery time in Singapore is between 1--2 months, it is likely that the Singapore society was still recovering at the end of the survey. This means that quantities measured in Time Period 1 would have different values to those measured in Time Period 2. Revisiting Table \ref{table:estimates}, we see that $C_2 = 12.00 \pm 1.27$ (the average number of contacts in Time Period 2) is slightly larger than $C_1 = 10.00 \pm 0.13$ (the average number of contacts in Time Period 1). In fact, participants inherited fewer contacts $C'_2 = 5.83 \pm 0.42$ in Time Period 2 compared to $C'_1 = 8.0 \pm 0.6$ in Time Period 1, but gained more new contacts $C''_2 = 4.0 \pm 0.1$ in Time Period 2 than $C''_1 = 0.0 \pm 0.0$ in Time Period 1. Interestingly, the numbers of contacts lost, $\Gamma_1 = 1.0 \pm 0.2$ and $\Gamma_2 = 1.0 \pm 0.4$, which do not require participants to do anything, are the same for Time Periods 1 and 2.

\begin{table}[htbp]
\tbl{Summary of stationarity statistics. In the last column, the count is less than 2,057 if no value was reported in Question B5 or Question B6.}{
\begin{tabular}{cccccc}
\hline
 & median & error & 95\% low & 95\% high & count \\
\hline
$S$ & $0.11$ & 0.02 & $0.07$ & $0.15$ & 2047 \\
$S'$ & $-0.22$ & $0.0$ & $-0.22$ & $-0.22$ & 1534 \\
$S''$ & $1.33$ & $0.09$ & $1.33$ & $1.6$ & 1722 \\
$S_g$ & $0.94$ & $0.1$ & $0.79$ & $1.2$ & 1538 \\
\hline
\end{tabular}}\label{table:stationarity}
\end{table}

We can quantify this non-stationarity more rigorously than just comparing the averages across the two three-month time periods. For this purpose, we compiled measurements in Table \ref{table:stationarity} for the quantities
\begin{align}
S &= \frac{2(C_2 - C_1)}{C_2 + C_1}, \\
S' &= \frac{2(C'_2 - C'_1)}{C'_2 + C'_1}, \\
S'' &= \frac{2(C''_2 - C''_1)}{C''_2 + C''_1}, \\
S_g &= \frac{2(\Gamma_2 - \Gamma_1)}{\Gamma_2 + \Gamma_1}
\end{align}
describing the fractional changes between measurements over the two time periods. As we can see, the total number of contacts $C_2$ in Time Period 2 increased by about $10\%$ from the total number of contacts $C_1$ in Time Period 1. In addition, the number of contacts $C'_2$ surviving from the previous time period decreased by 20$\%$ from $C'_1$, while the number of newly acquired contacts $C''_2$ in Time Period 2 increased by 133$\%$ over $C''_1$ acquired in Time Period 1. In contrast, the number of contacts lost $\Gamma_2$ in Time Period 2 increased by 94$\%$ over $\Gamma_1$ in Time Period 1, as seen in table \ref{table:stationarity}. Note that these measurements are performed on each node before the median value is determined. This explains why the median value of $S_g$ is not zero, which is what we obtain when we simply substitute the median values of $\Gamma_1 = 1$ and $\Gamma_2 = 1$ into the formula for $S_g$.

We can interpret these results to imply that consistent with ongoing social network recovery, people gained contacts during the six-month survey period. Likewise, as the contact gains increased, so did the contact losses. Both of these trends are visible in the time-evolving lockdown simulation, which shows that during the recovery phase gains exceed losses, but more subtly, the gain rate is decreasing while the loss rate is increasing until the two rates become equal in the steady state. We can compare these measures with the count of net gains, $R_0 + R_1 = 3.25$ and estimated net losses $2\Gamma = 2$ that also demonstrate non-stationarity due to ongoing network growth. Overall, this suggests that what we measured from the survey could be up to 50\% different from a survey that would be done later, after the Singapore society had reached steady state again. A related question whose answer we will never be able to get would be, what was the steady state of the Singapore social network like before the pandemic? This is important, as the COVID-19 pandemic may have permanently weakened the social fabric in Singapore. Alternatively, we could also have learnt from the experience and become more resilient than we were before.

Finally, let us say something about the dangers of using $N = 50$, $z^* = 5$ simulations to calibrate the mJGN model for Singapore society. Presumably, in future if we want `accurate' answers to pressing questions on the Singapore society we would run simulations with $N = 5,500,000$ nodes (still leaving aside the question on the heterogeneity of nodes). We have argued that this is the best we can do for now, and that the calibration results are still meaningful, because we took care to compare only intensive quantities. However, we also know from statistical thermodynamics that even intensive quantities are affected by finite system size. If we assume that the finite system size corrections are $O(1/N)$, the parameters $r_0$, $r_1$, and $\gamma$ estimated from the $N = 50$ simulations may be 1--10\% (depending on the constant prefactor of $1/N$) different from parameters estimated from the (hypothetical) $N = 5,500,000$ simulations. The standard way to eliminate these finite-size corrections is to run simulations for different system sizes, and thereafter extrapolate their parameters to $1/N \to 0$. At the very least, we would need simulations for one more system size, preferably for $N = 10^4$ nodes, which is a system size that is at the logarithmic midpoint between $N = 50$ and $N = 5,500,000$. If we can do such simulations, it would be possible to construct the $N \to \infty$ phase diagram of the mJGN model, and show the parameter point estimated for Singapore.

\section{Conclusions}
\label{sect:conclusions}

To conclude, we have in this paper demonstrated the use of phone contact data collected from a survey of $N = 2,057$ individuals living in Singapore (5,500,000 residents and non-residents) to calibrate a minimal dynamical social network model. When asked about their existing and new contacts over the past six months, survey participants reported $7.00 \pm 0.42$ contacts, as well as $2.00 \pm 0.21$ new random contacts and $1.25 \pm 0.11$ new social contacts over the six-month period. They also reported losing $2.0 \pm 0.6$ contacts over the six-month period. Using simulations of the model with $N = 50$ nodes with at most $z^* = 5$ connections, we mapped these survey results to parameter values $r_0 = 0.01$, $r_1 = 0.002$, $\gamma = 0.005$, and time step size of $\Delta T = 0.475$ day. We then simulated this best-fit one-community model to the Singapore society to find it recovering from a 16-month lockdown in 1--2 months. This suggested a fairly resilient Singapore society, even though we do not have similar estimates from other societies to compare against. By splitting the survey participants into two roughly equal groups, one with ages below or equal to 40, and the other with ages above 40, we estimated the probability for forming a new random contact between groups to be $\alpha = 0.305$. We also estimated the relative rate for forming new random contacts to be $\beta = 0.582$ in the young group, and $1 - \beta = 0.418$ in the old group.

From our analysis of the survey data, we also realized that there is more heterogeneity within the survey participants than what we expect from the one-community and two-community models. This led us to work with medians instead of means in our analysis. Our survey was carried out at a time when the Singapore society was still recovering from COVID-19 lockdowns, and therefore our survey results were not statistically stationary over time. Finally, we also acknowledged the large gap between the size of the Singapore society ($N = 5,500,000$) and the size of the model simulated ($N = 50$). These challenges suggest the following future works to more convincingly calibrate the model:
\begin{enumerate}
\item independently estimate the severity $\sigma$ of the lockdown;

\item repeat the survey during a socially stable period of time, after recovery is complete;

\item improve the survey questions:

    \begin{enumerate}
        \item modify QB6 so that the time period is also specified;

        \item modify questions so that we can estimate $R_0$, $R_1$, and $\Gamma$ independently for the two time periods;
        
    \end{enumerate}

\item modify QB11 to include more than five contacts, for instance up to $z = 7$;

\item modify two-community model to allow for directional asymmetry;

\item determine the phase diagrams of the one-community and two-community models;

\item apply the finite-size scaling of the one-community and two-community models.

\end{enumerate}

\section*{Acknowledgments}
This work is an outcome of the Future Resilient Systems project at the Singapore-ETH Centre (SEC) supported by the National Research Foundation, Prime Minister’s Office, Singapore under its Campus for Research Excellence and Technological Enterprise (CREATE) programme. Hans J. Herrmann thanks FUNCAP for financial support.

\appendix

\section{Survey Questions}
\label{app:surveyquestions}

Section A of the survey is related to demographics. The 13 questions, and the types of responses required, are shown below:

\begin{enumerate}
\item To which gender do you most identify?
    \begin{enumerate}[label=\SquareShadowBottomRight]
    \item Female
    \item Male
    \item Prefer not to say
    \end{enumerate}

\item How old are you? \underline{\hspace*{3cm}}

\item Which of the following ethnicities best describes you?
    \begin{enumerate}[label=\SquareShadowBottomRight]
        \item Chinese
        \item Malay
        \item Indian
        \item Eurasian
        \item Other, please specify: \underline{\hspace*{3cm}}
        \item Prefer not to say
    \end{enumerate}

\item What religion do you belong to or identify yourself most close to?
    \begin{enumerate}[label=\SquareShadowBottomRight]
        \item Buddhism
        \item Christianity
        \item Islam
        \item Taoism
        \item Hinduism
        \item Sikhism
        \item None
        \item Other, please specify: \underline{\hspace*{3cm}}
        \item Prefer not to say
    \end{enumerate}

\item What is your current marital status?
    \begin{enumerate}[label=\SquareShadowBottomRight]
        \item Single
        \item Married
        \item Widowed
        \item Separated/Divorced
        \item Other, please specify: \underline{\hspace*{3cm}}
        \item Prefer not to say
    \end{enumerate}

\item What is the highest educational level you have completed?
    \begin{enumerate}[label=\SquareShadowBottomRight]
        \item Primary \& below
        \item Secondary
        \item Nitec/Higher Nitec
        \item A levels/Diploma
        \item Bachelor's degree
        \item Master's degree
        \item Doctorate degree (PhD)
        \item Other, please specify: \underline{\hspace*{3cm}}
        \item Prefer not to say
    \end{enumerate}

\item What is your current employment status?
    \begin{enumerate}[label=\SquareShadowBottomRight]
        \item Employed (full/part time)
        \item Unemployed (seeking employment)
        \item Unemployed (not seeking employment)
        \item Retired
        \item Homemaker
        \item Other, please specify: \underline{\hspace*{3cm}}
        \item Prefer not to say
    \end{enumerate}

\item Over the past 12 months, what is the estimated average earnings (SGD) of your household per month?
    \begin{enumerate}[label=\SquareShadowBottomRight]
        \item Below \$2,000 per month
        \item Between \$2,000 and \$5,999
        \item Between \$6,000 and \$9,999
        \item \$10,000 and above
        \item Don't know
        \item Prefer not to say
    \end{enumerate}

\item What type of dwelling do you live in?
    \begin{enumerate}[label=\SquareShadowBottomRight]
        \item HDB 1-room
        \item HDB 2-room
        \item HDB 3-room
        \item HDB 4-room
        \item HDB 5-room or Executive flat
        \item Condominium or Private flat
        \item Landed property
        \item Other, please specify: \underline{\hspace*{3cm}}
        \item Prefer not to say
    \end{enumerate}

\item Is your dwelling owned or rented by yourself?
    \begin{enumerate}[label=\SquareShadowBottomRight]
        \item Owned/Co-owned
        \item Rented/Co-rented
        \item Other, please specify: \underline{\hspace*{3cm}}
        \item Prefer not to say
    \end{enumerate}

\item With whom are you currently living? Check all that apply.
    \begin{enumerate}[label=\SquareShadowBottomRight]
        \item Alone
        \item Live with spouse
        \item Live with children
        \item Live with grandchildren
        \item Living with partner
        \item Live with parents
        \item Live with other relatives
        \item Live with friend(s)
        \item Other, please specify: \underline{\hspace*{3cm}}
        \item Prefer not to say
    \end{enumerate}

\item How many children do you have? \underline{\hspace*{3cm}}

\item Do you work from home or at the office? \underline{\hspace*{3cm}}

\end{enumerate}

Section B of the survey is related to smart phone contacts. Through a few initial trials, we realized that instructions must be be provided. The instructions given to participants are thus as follows:
\begin{quote}
In all the following questions, when asked for a number, please provide your best estimate
(unless you know the exact answer), and answer with one single number only (do not key in
several numbers or ranges or words).
\end{quote}
The 13 questions, and the types of responses required, are shown below:

\begin{enumerate}

\item How many different contacts do you have in your smartphone’s contact list? \underline{\hspace*{3cm}}

\item Which of the following messaging apps do you use most often (in terms of time)?
    \begin{enumerate}[label=\SquareShadowBottomRight]
        \item WhatsApp
        \item Messenger
        \item Telegram
        \item Line
        \item Other, please specify: \underline{\hspace*{3cm}}
    \end{enumerate}

\item Which of the following messaging apps do you use to communicate with most people?
    \begin{enumerate}[label=\SquareShadowBottomRight]
        \item WhatsApp
        \item Messenger
        \item Telegram
        \item Line
        \item Other, please specify: \underline{\hspace*{3cm}}
    \end{enumerate}

\item How many different people did you exchange (receive and/or send) \emph{\textbf{at least one}} message with over the past 6 months? And over the past 3 months?
\emph{Please provide your best estimate.} \underline{\hspace*{3cm}}

\item Among those people with whom you have exchanged at least one message over the
past 6 months, how many had you \emph{never} exchanged messages with before?
\emph{Please provide your best estimate.} \underline{\hspace*{3cm}}

\item Among people with whom you have exchanged at least one message over the past 6
months, how many did you get to know through other contacts of yours?
\emph{Please provide your best estimate.} \underline{\hspace*{3cm}} \\
\emph{Examples: a friend of a friend, a dentist recommended by your doctor, an intern recommended by a colleague, a shop owner whose products have been advised to you by a friend, etc.}

\item Overall, did you use to make new contacts more often:
\emph{before} Covid, \emph{during} Covid, or \emph{about the same} before and during Covid?
    \begin{enumerate}
        \item Before Covid
        \item During Covid
        \item About the same
    \end{enumerate}

\item If your answer was 1 (more new contacts before Covid) or 2 (more new contacts during
Covid), how many times more would you estimate, approximately?
\emph{Please provide your best estimate.} About \underline{\hspace*{1cm}} times more.

\item How many people do you actively communicate with using messaging apps? \underline{\hspace*{3cm}}

\item Which app do you use most in the following contexts and/or groups of people?
    \begin{enumerate}[label=\SquareShadowBottomRight]
        \item Family: \underline{\hspace*{3cm}}
        \item Friends: \underline{\hspace*{3cm}}
        \item Information groups (government information, community activities…): \underline{\hspace*{3cm}}
        \item Business and professional activities: \underline{\hspace*{3cm}}
        \item Commercial activities (deliveries, advertisement, \dots): \underline{\hspace*{3cm}}
        \item Other, please specify: \underline{\hspace*{3cm}}
    \end{enumerate}

\item Think about the 5 persons with whom you communicate most regularly:
    \begin{enumerate}
        \item How old are they?
            \begin{enumerate}
                \item Person 1: \underline{\hspace*{3cm}}
                \item Person 2: \underline{\hspace*{3cm}}
                \item Person 3: \underline{\hspace*{3cm}}
                \item Person 4: \underline{\hspace*{3cm}}
                \item Person 5: \underline{\hspace*{3cm}}
            \end{enumerate}
        \item What is their gender?
            \begin{enumerate}
                \item Person 1: \underline{\hspace*{3cm}}
                \item Person 2: \underline{\hspace*{3cm}}
                \item Person 3: \underline{\hspace*{3cm}}
                \item Person 4: \underline{\hspace*{3cm}}
                \item Person 5: \underline{\hspace*{3cm}}
            \end{enumerate}
        \item Which ethnicity best describes them?
            \begin{enumerate}
                \item Person 1: \underline{\hspace*{3cm}}
                \item Person 2: \underline{\hspace*{3cm}}
                \item Person 3: \underline{\hspace*{3cm}}
                \item Person 4: \underline{\hspace*{3cm}}
                \item Person 5: \underline{\hspace*{3cm}}
            \end{enumerate}
        \item What is the highest educational level they have completed?
            \begin{enumerate}
                \item Person 1: \underline{\hspace*{3cm}}
                \item Person 2: \underline{\hspace*{3cm}}
                \item Person 3: \underline{\hspace*{3cm}}
                \item Person 4: \underline{\hspace*{3cm}}
                \item Person 5: \underline{\hspace*{3cm}}
            \end{enumerate}
        \item Do they work in the same industry as you?
            \begin{enumerate}
                \item Person 1: \underline{\hspace*{3cm}}
                \item Person 2: \underline{\hspace*{3cm}}
                \item Person 3: \underline{\hspace*{3cm}}
                \item Person 4: \underline{\hspace*{3cm}}
                \item Person 5: \underline{\hspace*{3cm}}
            \end{enumerate}
        \item What religion do they belong to or identify themselves closest to?
            \begin{enumerate}
                \item Person 1: \underline{\hspace*{3cm}}
                \item Person 2: \underline{\hspace*{3cm}}
                \item Person 3: \underline{\hspace*{3cm}}
                \item Person 4: \underline{\hspace*{3cm}}
                \item Person 5: \underline{\hspace*{3cm}}
            \end{enumerate}
        \item What is their current marital status?
            \begin{enumerate}
                \item Person 1: \underline{\hspace*{3cm}}
                \item Person 2: \underline{\hspace*{3cm}}
                \item Person 3: \underline{\hspace*{3cm}}
                \item Person 4: \underline{\hspace*{3cm}}
                \item Person 5: \underline{\hspace*{3cm}}
            \end{enumerate}
    \end{enumerate}

\item There are some people that you used to communicate with 3 to 6 months ago, and with
whom you stopped communicating with since then. Would you say they were mainly:
    \begin{enumerate}[label=\SquareShadowBottomRight]
        \item Friends
        \item Colleagues or related professional partners
        \item Related to information groups (government information, community activities…)
        \item Related to commercial activities (deliveries, advertisement…)
        \item Other, please specify: \underline{\hspace*{3cm}}
    \end{enumerate}

\item What is the main reason for you to stop communicating with someone?
    \begin{enumerate}[label=\SquareShadowBottomRight]
        \item Lost interest in the person
        \item Having an argument
        \item Social distancing rules
        \item Physical distance (e.g., living in different countries)
        \item Commercial or business interest has ended
        \item Other, please specify: \underline{\hspace*{3cm}}
    \end{enumerate}

\end{enumerate}


\bibliographystyle{ws-ijmpc}
\bibliography{main.bib}

\end{document}